\documentclass[twocolumn]{article}
\usepackage{abstract}
\usepackage[top=2cm, bottom=3cm, left=2cm, right=2cm]{geometry}
\let\oldbibliography\thebibliography
\renewcommand{\thebibliography}[1]{%
  \oldbibliography{#1}%
  \setlength{\itemsep}{0pt}%
}
\usepackage{graphicx,xcolor}
\usepackage[fleqn]{amsmath}
\usepackage{newtxtext}
\usepackage[varg]{newtxmath}
\usepackage{array}
\newcommand{\corr}[1]{#1}
\newcommand{\corri}{white}
\def\dd{{\mathrm{d}}}
\def\diver{\mathop{\mathrm{div}}}
\def\curl{\mathop{\mathrm{curl}}}

\def\grad{\mathop{\mathrm{grad}}}
\def\sub#1{_{\mbox{\scriptsize#1}}}
\def\sur#1{^{\mbox{\scriptsize#1}}}
\def\vct#1{{\mathchoice{\mbox{\boldmath$#1$}}{\mbox{\boldmath$#1$}}%
  {\mbox{\scriptsize\boldmath$#1$}}{\mbox{\scriptsize\boldmath$#1$}}}}
\newcommand{\pdfrac}[2]{\frac{\partial#1}{\partial#2}}
\newcommand{\vnabla}{\vct{\nabla}}
\newcommand{\NN}{\nonumber\\}
\newcommand{\dt}[1]{\frac{\dd#1}{\dd t}}
\newcommand{\incr}{\Delta}

\title{
Why the controversy over displacement currents never ends?
}
\author{
Masao Kitano
}
\date{
{\small\it
Center for Quantum Information and Quantum Biology,
Osaka University, Osaka 560-0043, Japan
\\[4pt]
Research Institule of Applied Science,
Sakyoku, Kyoto 606-8202, Japan
\\
Professor emeritus, Kyoto University,
Kyoto 606-8501, Japan
}
\\[2ex]
\small
\today
}

\begin{document}
\twocolumn[
\maketitle
\begin{onecolabstract}
\noindent
Displacement current is the last piece of the puzzle of electromagnetic theory.
Its existence implies that electromagnetic disturbance can propagate at the speed of light and finally it led to the discovery of Hertzian waves.
On the other hand, since magnetic fields can be calculated only with conduction currents using Biot-Savart's law,
a popular belief that displacement current does not produce magnetic fields
has started to circulate.
But some people think if this is correct, what is the displacement current introduced for.
The controversy over the meaning of displacement currents has been going on for more than hundred years.
Such confusion is caused by forgetting the fact that in the case of non-stationary currents, neither magnetic fields created by conduction currents nor those created by displacement currents can be defined.
It is also forgotten that the effect of displacement current is automatically incorporated in the magnetic field calculated by Biot-Savart's law.
In this paper,
mainly with the help of Helmholtz decomposition,
we would like to clarify the confusion surrounding displacement currents and
provide an opportunity to end the long standing controversy.
\\
\textit{Keywords:} displacement current, Biot-Savart law, Ampere's law, Maxwell-Ampere's law, Helmholtz's decomposition, non-stationary current
\end{onecolabstract}
\vspace{\baselineskip}
]

\section{Introduction}
The time derivative of the electric flux density,
$\partial_t\vct D$, is named
displacement current density,
which is the final piece to complete the hard puzzle
of electromagnetic theory.
This discovery made by James Clerk Maxwell
\cite{maxwell-paper,simpson} was possible
only through his keen eyes
forseeing its existence from theoretical inevitability
(1864).

He found the fact that
the propagation
velocity of the wave solution enabled by the displacement current,
was consistent with the speed of light,
which was already measured experimentally at that time.
The value of constant $(\mu_0\varepsilon_0)^{-1}$
had been determined by Weber and Kohlrausch
in other context\cite{weber-kohlrausch},
where $\mu_0$ and $\varepsilon_0$ are the permeability and
permittivity of vacuum, respectively.
Maxwell was convinced that light is an electric and magnetic disturbance
propagating in a vacuum.
Later, H.R. Hertz discovered radio waves (1888)
in attempting to detect displacement currents using a capacitor.

Displacement currents occupy an important position in electromagnetics.
However,
owing to the fact that magnetic fields can be correctly calculated by the Biot-Savart law,
which does not seem to include the displacement current,
it has widely been claimed that
{\em the displacement
current does not produce a magnetic field}.
As a matter of fact, however,
the Biot-Savart law implicitely includes the contribution of displacement currents.

In this paper, we would like to clarify the confusion surrounding displacement currents and
its causes
and help to promote correct understanding.

\section{Magnetic Action of Electric Currents
and Displacement Current}

In 1820, Hans Christian {\O}rsted discovered that
a compass needle swings
in response to an electric current flowing near it.
That same year, the relation between current and magnetic field was formulated in two ways;
Biot-Savart's law and Amp\`{e}re's law,
which correspond to Coulomb's law
and
Gauss's law in electrostatics, respectively.

These magnetic field laws were based on the assumption that the current is flowing through a closed circuit.
Almost half a century later,
considering the case of unclosed current,
as in the case of charging capacitor,
Maxwell theoretically derived the necessity of
displacement currents.

His argument goes as follows.
By taking the divergence of both sides of
Amp\`{e}re's equation,
$
  \curl\vct H = \vct J
$,
we have,
$
  0 = \diver\vct J
$,
with the identity $\diver\curl = 0$.
In other words, Amp\`{e}re's equation implicitely
assumes divergence-free currents.
This is also called the "steady-state current condition,"
because
the charge conservation law,
$\partial_t\varrho = -\diver\vct J$,
implies steady charge distributions.
($\partial_t = \partial/\partial t$ is used for brevity.)
If this condition is not satisfied, i.e., $\diver\corr{\vct J}\neq0$,
Amp\`{e}re's law must
be modified as follows:
\begin{align}
    \curl\vct H = \vct J + \partial_t\vct D
.
\label{eq:maxwell-ampere}
\end{align}
The displacement current density term, $\partial_t\vct D$ is added.
Now, taking the divergence of both sides, we have
\begin{align*}
  0 = \diver\vct J + \partial_t\diver\vct D =
  \diver\vct J + \partial_t\varrho
  .
\end{align*}
The time derivative of Gauss's formula, $\diver\vct D=\varrho$, is used.
This is consistent with the charge conservation law.

Based on this reasoning,
in his treatise \cite{maxwell-treatise}, Maxwell states
\begin{quote}
One of the chief pecurialities of this treatise is the
doctrine which it asserts, that the true electric current
$\mathfrak C$ ($\vct C$), that
on which the electromagnetic phenomena depend, is not the
same thing as $\mathfrak K$ ($\vct J$), the current of conduction,
but that the time
variation of $\mathfrak D$ ($\vct D$), the electric displacement, must be taken into
account in estimating the total movement of electricity, so that
we must write,
\begin{align*}
\mathfrak C = \mathfrak K + \dot{\mathfrak D},\quad
\text{\footnotesize (Equation of True Currents)}
.
\end{align*}
\end{quote}
Hereafter we write the true (total) current as
\begin{align*}
  \vct J\sub{tot} = \vct J + \partial_t\vct D
.
\end{align*}

Equation (\ref{eq:maxwell-ampere}) is now called
Maxwell-Amp\`{e}re's equation.
Maxwell's electromagnetic theory thus created is being organized
by the followers and spread to the academic world

Oddly, \corr{however}, the doctrine
that displacement currents do not create magnetic fields
began to circulate.
The main reasons are
\begin{itemize}
\item
Even in the presence of displacement currents,
magnetic field is calculated correctly by the Biot-Savart equation.
\item
A typical displacement current is one that occurs where the linear current is interrupted.
Then charges are accumulated at the endpoint and yield a spherically
symmetric electric field.
The corresponding displacement current is also spherically symmetric
\corr{and} the resultant magnetic field
vanishes.
\end{itemize}

Various arguments against, for, or from a neutral standpoint about this theory,
some of which seem to deepen the confusion, are continuing
in papers and textbooks
\cite{warburton,rosser,planck,arthur,selvan,%
weber,wolsky,
bierman,french,roche1,jackson1,roche2,jackson2,purcell}.

In this paper, we will show
the claim that displacement current does not create a magnetic field is
due to a lack of understanding of the mathematical structure of electromagnetic fields.
We mainly discuss from the following points of view:
\begin{itemize}
  \item
  It is impossible in principle to separate magnetic fields into those caused by
  "conduction currents" and those caused by "displacement currents".
  \item
  The posed question "does a displacement current create a magnetic field or not?"
  is logically meaningless.
  \item
  Contrary to popular perception,
  the Bio-Savart law perfectly includes the effect of displacement currents implicitely \cite{griffiths-book}.
\end{itemize}

\section{Need for Displacement Current}

In this section, we will reconfirm how
the Amp\`{e}re's law is modified to account for
displacement currents.

The integral form of Amp\`{e}re's law, $\curl\vct H=\vct J$, is
\begin{align}
  \int_C\vct H\cdot\dd\vct l = \int_{S}\vct J\cdot\dd\vct S
  ,
\label{eq:ampere}
\end{align}
where
the closed path $C=\partial S$ is the edge of the surface $S$.
In this equality
the surface $S$ can be arbitrary as long as the closed path $C$ is its edge.
In order for the integral to be the same regardless of the surface,
$\diver\vct J = 0$ must be satisfied everywhere
(steady-state current condition).
Otherwise, for surfaces $S_1\neq S_2$ with $\partial S_1=\partial S_2=C$,
the divergence theorem gives
\begin{align*}
  0\neq \int_V \diver\vct J\,\dd v
  = \left(\int_{S_1}-\int_{S_2}\right)\vct J\cdot\dd\vct S
,
\end{align*}
where $V$ is the volume enclosed
by $S_1$ and $S_2$.

\begin{figure}[t]
  \centering
\fcolorbox{\corri}{white}{
 \includegraphics[scale=0.38]{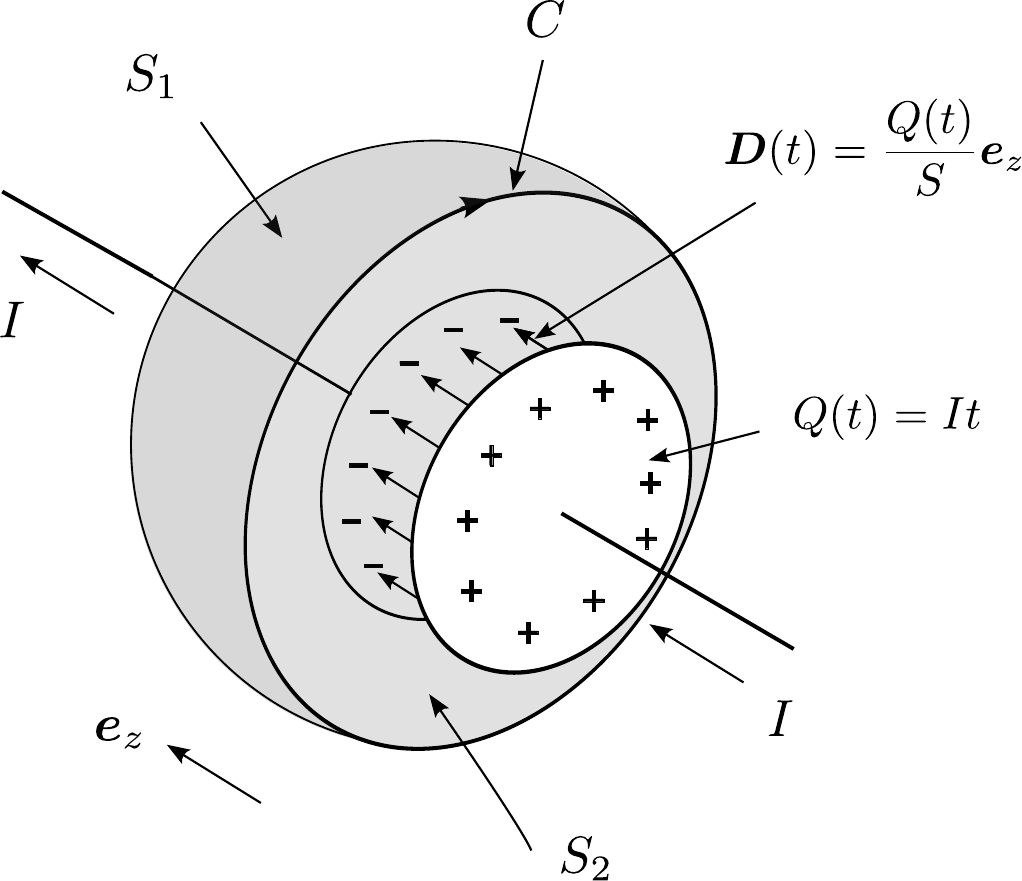}
}
  \caption{Displacement currents in charging capacitor}
  \label{fig:capacitor}
\end{figure}

Consider a capacitor
being charged with a constant current $I$, as shown in Fig \ref{fig:capacitor}.
\corr{We have two surfaces $S_1$ and $S_2$
that share the same circle $C$
encircling the capacitor
as their respective circumferences.
While the hemisphere $S_1$ crosses the current $I$,
the disk $S_2$ passes between the electrodes of the capacitor
and crosses no currents.
}
The integrals of the current density for these surfaces
\begin{align*}
  I = \int_{\corr{S_1}}\vct J\cdot\dd\vct S \quad\neq\quad
  \int_{\corr{S_2}}\vct J\cdot\dd\vct S = 0
,
\end{align*}
are clearly not equal.
But if we add
the displacement current
density $\partial\vct D/\partial t$,
then the surface integral for \corr{$S_2$} becomes
\begin{align*}
  \int_{\corr{S_2}}\left(\vct J + \pdfrac{\vct D}{t}\right)\cdot\dd\vct S =
  \int_{\corr{S_2}}\pdfrac{\vct D}{t}\cdot\dd\vct S =
  \dt{}Q(t) = I
  ,
\end{align*}
and now the equality holds.
The charge on the capacitor plate
$Q = \int_{\corr{S_2}}\vct D\cdot\dd\vct S$
can be derived from $\vct D$ between the plates.
With this model we can confirm that
\corr{Eq.~(\ref{eq:ampere})} must be modified as
\begin{align}
  \int_C\vct H\cdot\dd\vct l = \int_{S} \left(\vct J +\pdfrac{\vct D}{t}\right)\cdot
  \dd\vct S
  .
\label{eq:ampere-int}
\end{align}
This is the integral form of Maxwell-Amp\`{e}re's equation.

\begin{figure*}[t]
\centering
\includegraphics[scale=0.48]{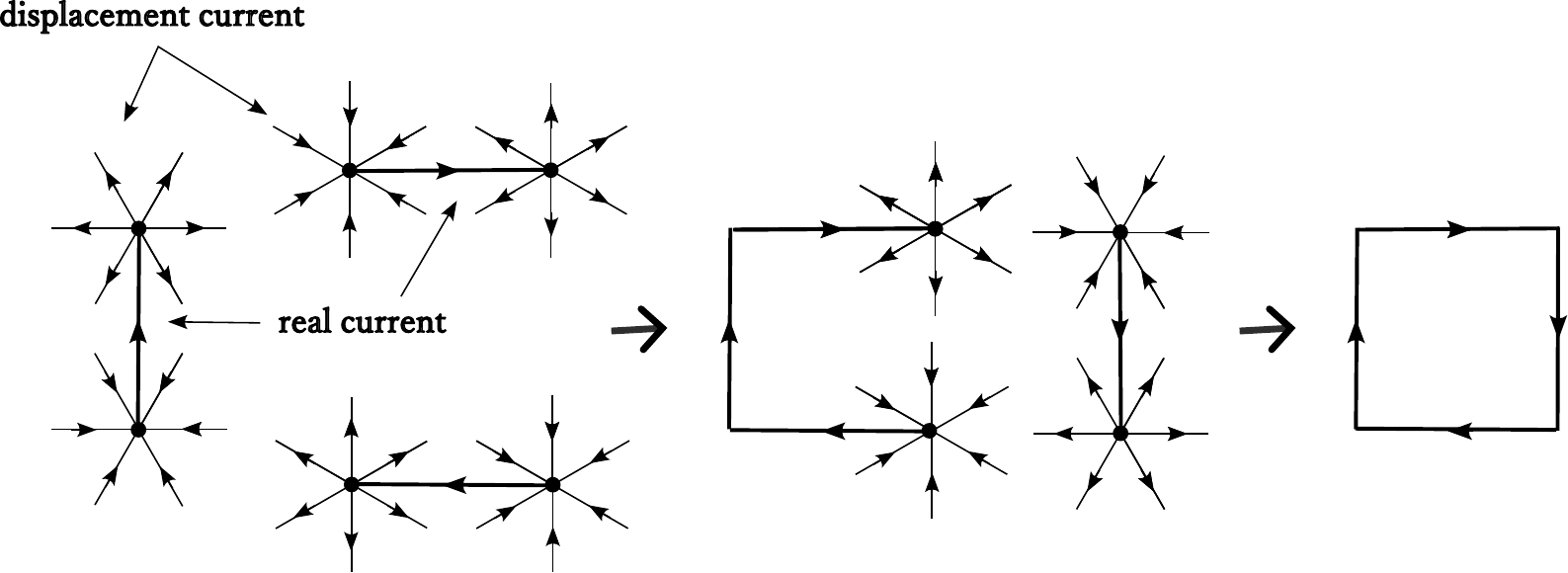}
\caption{Counteracting displacement currents in current element connections}
\label{fig:biotsavart}
\end{figure*}

\section{Inappropriate Problem Setting}
In this section, we show that
the question whether the displacement current creates a magnetic field or not
is totally meaningless.

The question can be broken down
in the following way.
When the total current density $\vct J\sub{tot}$
is divided
into the conduction current density
$\vct J$ and the displacement current density $\partial_t\vct D$,
the magnetic field can be divided into two components as
$\vct H = \vct H\sub c + \vct H\sub d$,
correspondingly.
And if we can prove $\vct H\sub d=0\,(\neq0)$,
then the answer is no (yes).
But, in the first place, what are the (local) equations that these magnetic fields obey?
The only possible choice seems to be
\begin{align}
      \curl\vct H\sub c \stackrel{\corr{?}}{=} \vct J,\quad
      \curl\vct H\sub d \stackrel{\corr{?}}{=} \partial_t \vct D
,
\label{eq:wrong-division}
\end{align}
but it
leads us to the contradiction when we take the divergence;
\begin{align*}
   0 \stackrel{\corr{?}}{=} \diver\vct J\neq 0
   ,\quad
   0 \stackrel{\corr{?}}{=} \diver\partial_t\vct D\neq 0
   .
\end{align*}
What did we introduce the displacement current for?

We have found that
{\em the magnetic field
created the displacement current cannot be defined} and therefore,
it makes no sense to ask whether such
an undefinable quantity is zero or not.

In general, if we want to solve the
Maxwell-Amp\`{e}re equation,
$\curl\vct H = \vct J + \partial_t\vct D$,
by superposition, then we should set
\begin{align}
\curl\vct H_1 = \vct J_1 + \partial_t\vct D_1
,\quad
\curl\vct H_2 &= \vct J_2 + \partial_t\vct D_2
,
\label{eq:proper-division}
\end{align}
and divide not only the current density but also the displacement current density term, so that each of them satisfies
\begin{align}
\diver(\vct J_1 + \partial_t\vct D_1) = 0,
\quad \diver(\vct J_2 + \partial_t\vct D_2) = 0
.
\end{align}
This is the
{\em proper division of total current}.
On the other hand
the division of
Eq. (\ref{eq:wrong-division}) makes no sense.

Let us consider a mathematical case to see why a simple-minded superposition
does not hold.
For a general linear map $A:\, X\rightarrow Y$, from a space $X$ to another $Y$, let
$
  \text{Im}\,A = \{ A\vct x\mid \vct x\in X\} \subset Y,
$
which is called the range or image of $A$.
In order that the linear equation
\begin{align*}
A\vct x = \vct b
,
\end{align*}
has a solution $\vct x\in X$, the condition
$\vct b\in \text{Im}\,A$ must be met.
Even when
$
\vct b = \vct b_1 + \vct b_2 \in \text{Im}\,A
$,
if
$
\vct b_1,\;\vct b_2 \notin \text{Im}A
$
then the superposition cannot be used.
Because
\begin{align*}
  A\vct x_1 = \vct b_1,\quad
  A\vct x_2 = \vct b_2
  ,
\end{align*}
have no solutions.
An example is
\begin{align*}
  A=\begin{bmatrix}
    1 & 1 \\ 0 & 0
  \end{bmatrix}
,\quad
\vct b =
\begin{bmatrix}
  2 \\ 0
\end{bmatrix}
,\quad
\vct b_1 =
\begin{bmatrix}
  1 \\ 1
\end{bmatrix}
,\quad
\vct b_2 =
\begin{bmatrix}
  1 \\ -1
\end{bmatrix}
.
\end{align*}
\corr{
To make a situation where the solution is unique,
we should set up another linear equation, $B\vct x=0$,
which corresponds to $\diver(\mu_0\vct H)=0$ in our case.}

\section{Biot-Savart's Law and Displacement Current}
\label{sec:biotsavart}
Here we will show that
contrary to common belief
Biot-Savart's law
include the effect of displacement currents
\cite{jackson,bierman}.

For the sake of brevity,
the Coulomb field is written as
\begin{align*}
  \vct G(\vct r) := \frac{\vct r}{4\pi |\vct r|^3}
  .
\end{align*}
With this, we have
$
  \vnabla\cdot\vct G(\vct r) = \delta^3(\vct r)
$
,
$
  \vnabla\times\vct G(\vct r) = 0
$
,
$
 \vnabla(1/r) = -4\pi\vct G(\vct r)
$,
where $\delta^3(\vct r)$ is the delta function.
\corr{
(For calculation, we use $\vnabla\cdot$, $\vnabla\times$,
and $\vnabla$,
instead of $\diver$, $\curl$, and $\grad$.)}

The electric flux density for a charge $q$ placed at the origin is
$
  \vct D_q(\vct r) = q\vct G(\vct r)
$
and that for an electric dipole $\vct p = q\vct l$ at the origin is
\begin{align}
  \vct D_{\vct p}(\vct r) = -(\vct p\cdot\vnabla)\vct G(\vct r)
  .
\label{eq:dipole-electric-flux}
\end{align}
For the current element $I\incr\vct l$
placed at the origin,
the magnetic field created at the point $\vct r$ is
given by Biot-Savart law in the difference form
\begin{align}
\incr\vct H(\vct r) = I\incr\vct l\times
\vct G(\vct r)
.
\label{eq:magnetic-field}
\end{align}
The magnetic field due to the current $I$ flowing through
the closed circuit $L$
is given as a superposition (integral)
\begin{align}
  \vct H(\vct r)
  = \oint_L \dd\vct H(\vct r - \vct r')
  = \oint_L I\dd\vct l'\! \times\! \vct G(\vct r - \vct r')
  ,
\label{eq:biot-savart}
\end{align}
where
$\dd\vct l'$ is a line element at position $\vct r'$ on path $L$.
Originally, the condition
that "$L$ is closed"
was required by Biot-Savart's integrated formula initially.

Let us find the vortex of the magnetic field element (\ref{eq:magnetic-field}).
\corr{With
the help of a vector analysis formula,}
we have
\begin{align}
&\vnabla\times\incr\vct H(\vct r)
= \vnabla\times\left[(I\incr\vct l)\times\vct G(\vct r)\right]
\NN
&\hspace{3em}
= I\incr\vct l \left[\vnabla\cdot\vct G(\vct r)\right]
  - (I\incr\vct l\cdot\vnabla)\vct G(\vct r)
\NN
&\hspace{3em}
= I\incr\vct l\,\delta^3(\vct r)
   + {\pdfrac{}{t} \vct D_{It\incr\vct l}(\vct r)}
  =: \incr\vct J\sub{tot}(\vct r)
  .
\label{eq:current-element}
\end{align}
We note that
in addition to the original current element $I\incr\vct l$
the additional term appears.
This term is the time derivative of the electric flux density
(\ref{eq:dipole-electric-flux}) for the electric dipole
 $\vct p(t)=It\Delta\vct l$
at the origin.
It means that
if a constant current $I$ flows on the line element $\incr\vct l$, then
there accumulates charge $\pm Q(t)=\pm It$ at each end $\pm\incr\vct l/2$ to form
an electric dipole.

The total current
$\incr\vct J\sub{tot}$
(\ref{eq:current-element})
satisfies the stationary condition and
the magnetic field $\incr\vct H$ is generated by this total current $\incr\vct J\sub{tot}$.

The condition that
$L$ is closed, which was originally assumed in the Biot-Savart equation,
is actually unnecessary.
When a current is formally
integrated for an open path $L$ with start (end) point $\vct r_1$ ($\vct r_2$), we obtain
\begin{align}
\vct J\sub{tot}(\vct r) &=
\int_L\dd\vct J\sub{tot}(\vct r \! -\! \vct r')
\NN
&=
I\int_L\dd\vct l'\delta^3(\vct r \! - \! \vct r')
+ \left.
\rule{0mm}{2.5ex}
I\vct G(\vct r \! -\! \vct r') \right|_{\vct r'=\vct r_1}^{\vct r_2}
.
\end{align}
As shown in Fig.~\ref{fig:biotsavart},
when the current elements are connected (integrated),
the displacement currents from opposing endpoints cancel each other and
only those at the two extreme ends of the path remain.
This is similar to the case where small bar magnets are connected
to form a long chain.
The magnetic fields of opposite polarity at the connection points cancel each other, and only the magnetic fields from the poles at both ends survive.

In the end, the Biot-Savart's equation (\ref{eq:biot-savart})
can be applied even for unsteady (open) currents,
just with the modification of integration path;
$\oint\rightarrow \int$,
which is beyond the originally intended scope of application.

Although it is an integral of the current distribution $\vct J$,
the total current $\vct J\sub{tot} = \vct J + \vct J\sub{disp}$
is automatically taken into account and the corresponding magnetic field is given.

This fact has been pointed out by
from time to time
\cite{griffiths-book,milsom},
but
many people still mistakenly believe that
when using Biot-Savart
they can calculate
the magnetic field only
due to the conduction currents that they give.
But even if you don't order the displacment current, it will always come by itself.
This "hidden trick" is one of the sources
of confusion over the displacement current.

Even though both Biot-Savart's law and Amp\`{e}re's law were established
in the same year (1920),
only the former consealed the effect of displacement current that will be exposed
45 years later.

\begin{figure}[t]
  \centering
  \includegraphics[scale=0.30]{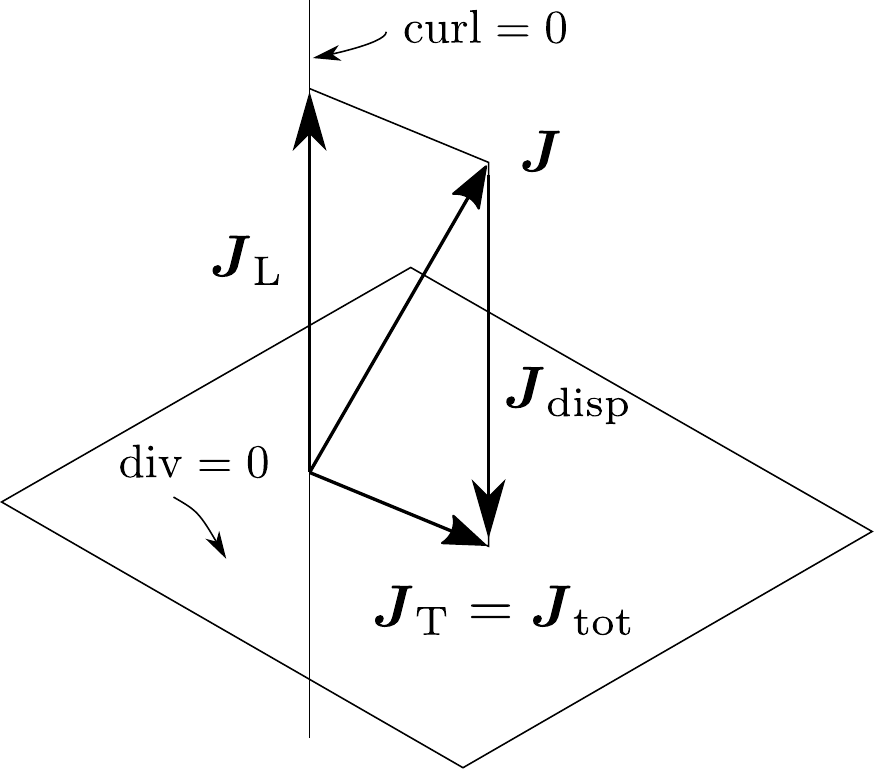}
  \caption{Helmhortz decomposition of current distribution}
  \label{fig:helmholz-decomposition}
\end{figure}

\begin{figure*}[t].
  \centering
  \begin{tabular}{m{8.0em}ccc}
  \hline
  & $\vct J\hspace{2em}=$ & $\hspace{2em}\vct J\sub{T}\hspace{3em}+$ & $\vct J\sub{L}$
  \\[0.4ex]
  & $\vct J\hspace{2em}=$ & $\hspace{2em}\vct J\sub{tot}\hspace{2.6em}+$ & $(-\vct J\sub{disp})$
  \\
  \hline
\rule{0mm}{1.8em}Current element
    &
  $\vct p\,\delta^3(\vct r)$
  &
  $(\vct p\times\vnabla)\times\vct G(\vct r)$
  &
  $(\vct p\cdot\vnabla)\vct G(\vct r)$
  \\
  Semi-infinite linear current &.
  \begin{minipage}{9em}
    \centering
  \includegraphics[scale=0.5]{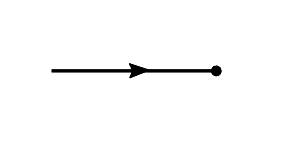}
  \end{minipage}
  &
  \begin{minipage}{9em}
  \includegraphics[scale=0.5]{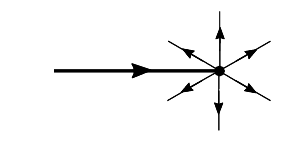}
  \end{minipage}
  &
  \begin{minipage}{9em}
  \includegraphics[scale=0.5]{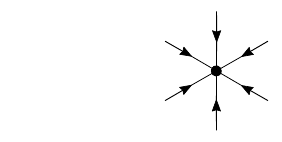}
  \end{minipage}
\\
Spherically symmetric current &.
  \begin{minipage}{9em}
    \centering
  \includegraphics[scale=0.5]{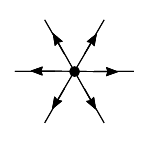}
  \end{minipage}
  &
  \begin{minipage}{9em}
    \centering
  $0$
  \end{minipage}
  &
  \begin{minipage}{9em}
    \centering
  \includegraphics[scale=0.5]{./spherical.pdf}
  \end{minipage}
\\
Charging capacitor&
  \begin{minipage}{9.2em}
  \centering
  \includegraphics[scale=0.6]{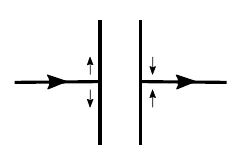}
  \end{minipage}
  &
  \begin{minipage}{9.2em}
  \centering
  \includegraphics[scale=0.6]{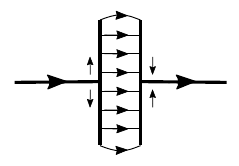}
  \end{minipage}
  &
  \begin{minipage}{9.2em}
  \centering
  \includegraphics[scale=0.6]{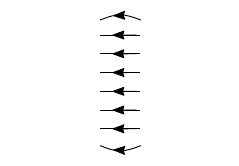}
  \end{minipage}
\\
  \hline
  \end{tabular}
  \caption{Examples of Helmholtz decomposition of
  current distribution}
  \label{fig:examples}
\end{figure*}

\section{Helmholtz Decomposition of Vector Fields}
From a mathematical point of view,
we take a closer look at
how the Biot-Savart equation automatically incorporates
the effect of displacement currents.
The current density field $\vct J$ (or a three dimensional
vector field in general)
can be divided into
the "curl-free" component $\vct J\sub{L}$ and the "divergence-free" component $\vct J\sub{T}$, namely,
\begin{align}
    &\vct J(\vct r) = \vct J\sub{L}(\vct r) + \vct J\sub{T}(\vct r),
    \NN
    &\hspace{1em}\curl\vct J\sub{L}(\vct r) = 0,
    \quad \diver\vct J\sub{T}(\vct r) = 0
    .
\label{eq:helmholtz-decomposition}
\end{align}
This is the so-called Helmholtz decomposition
\corr{\cite{griffiths-book,arfken,panofsky}}.
\corr{The uniqueness of the decomposition requires that the field quantities converge quickly to zero at infinity, which is satisfied in the present case.
}
The curl-free and divergence-free fields are also called
the ``longitudinal'' and ``transverse'' fields,
respectively, which
are denoted by subscripts ``L'' and ``T''.
The latter names are derived from the relations
of their Fourier transform to the wavevector $\vct k$;
\begin{align*}
\vct k\times(\mathcal F\vct J\sub{L})(\vct k)=0,\quad
\vct k\cdot(\mathcal F\vct J\sub{T})(\vct k)=0
.
\end{align*}

The Biot-Savart law
for the three-dimensional current density $\vct J(\vct r)$ is
\begin{align}
\vct H(\vct r)
= \int_V\! \dd v' \vct J(\vct r')\times\vct G(\vct r - \vct r')
,
\label{eq:biot-savart-3d}
\end{align}
and its vortex is
\begin{align}
\curl\vct H(\vct r)
=
\vct J(\vct r)
- \int_V \! \dd v'
(\vnabla'{\cdot}\vct J(\vct r'))\vct G(\vct r\! - \!\vct r')
.
\label{eq:curl}
\end{align}
The first term is just the given current density.
With the charge conservation of law
$\vnabla\cdot\vct J(\vct r) + \partial_t\varrho(\vct r, t) = 0$,
the second term turns out to be the displacement current density:
\begin{align*}
\int_V\! \dd v' \partial_t\varrho(\vct r', t)\vct G(\vct r - \vct r')
= \partial_t\vct D(\vct r, t)
.
\end{align*}
We can verify that $\partial_t\vct D$ is
longitudinal.

Equation (\ref{eq:curl}) can be rewritten as
\begin{align*}
\curl\vct H\corr{(\vct r)} = \vct J(\vct r) - \hat{L}\vct J(\vct r)
= (\hat 1 - \hat{L})\vct J(\vct r)
,
\end{align*}
where $\hat{1}$ is the identity operator, and
the operator
$\hat{L}$ acts on the vector field $\vct J(\vct r)$ to
create a new vector field:
\begin{align}
(\hat L\vct J)(\vct r)
&:= \int_{\corr{V}}\! \dd v' (\vnabla'{\cdot}\vct J(\vct r'))\vct G(\vct r - \vct r')
.
\end{align}
The operator $\hat{L}$ gives the longitudinal field components
of a vector field.

We define another operator
$\hat{T} := \hat 1 - \hat{L}$,
which gives the transverse field component
$\hat{T}\vct J$.
The operators
$\hat{L}$ and $\hat{T}$ are equipped with
the properties of projection operators, namely,
\begin{align*}
\hat{T}^2 = \hat{T},\quad
\hat{L}^2 = \hat{L},\quad
\hat{T}\hat{L} = \hat{L}\hat{T} = 0
.
\end{align*}

As shown in Fig.~\ref{fig:helmholz-decomposition}, the Biot-Savart law
gives the magnetic field due to
the transverse component of the
current density $\vct J\sub{T} =
\hat{T}\vct J = \vct J - \vct J\sub{L}$,
or due to the total current density
\begin{align*}
\vct J\sub{tot} := \vct J + \vct J\sub{disp}
.
\end{align*}
From these equations,
we know that the displacement current is
the curl-free (longitudinal) component with sign changed of the given current:
\begin{align}
\vct J\sub{disp} = -\vct J\sub{L} \: (=\partial_t\vct D)
.
\end{align}
Adding the displacement current density $\vct J\sub{disp}$
to the current density $\vct J$ means the cancellation of the longitudinal component $\vct J\sub{L}$
to yield $\vct J\sub{T}$ or $\vct J\sub{tot}$.
The difference between the subtraction and the sum gives
a very different impression.

In the cource of deriving
the Biot-Savart's law from Maxwell's
equations, we confirm why the former includes
the effect of displacement currents.
With the curl of the Maxwell-Amp\`{e}res's law,
$\curl\curl\vct H = \curl\vct J\sub T$,
$\vct J\sub T=\vct J \corr{\,+\,} \partial_t\vct D$,
and $\diver(\mu_0\vct H) = 0$, we have
\begin{align*}
  \nabla^2\vct H = -\curl\vct J\sub T
  ,
\end{align*}
where $\curl\curl=\grad\diver-\nabla^2$
is used.
The solution to this (vector) Poisson's equation
\cite{panofsky}
is
\begin{align}
  \vct H(\vct r)
  = \int\dd v'\vct J\sub{T}(\vct r')\times
  \vct G(\vct r - \vct r')
  =: (\hat{f}\sub{T}\vct J\sub T)(\vct r)
,
\end{align}
where the operator $\hat{f}\sub{T}$ is defined
so as to map a transverse current $\vct J\sub{T}$
to the corresponding magnetic field $\vct H$.
This map,
which can be written symbolically
$\vct H = \curl^{-1}\vct J\sub T$, is
invertible, i.e., one-to-one.
For a general current $\vct J$, we should have
\begin{align*}
\vct H = \hat{f}\sub{T}\corr{\hat{T}}\vct J
=: \hat{f}\sub{BS}\vct J
,
\end{align*}
which correponds to
the Biot-Savart's law (\ref{eq:biot-savart-3d}).
The two-step operation,
$\hat{f}\sub{BS} = \hat{f}\sub{T}\hat{L}$
is {\em not invertible}.
The current distribution cannot be
uniquely determined from the magnetic field.
This fact is overlooked so often.

In order to fix the problem
of improper division of Eq. (\ref{eq:wrong-division}),
we can apply $\hat{T}$ for each current.
\begin{align*}
  \curl\vct H = \hat{T}\vct J = \vct  J
  + \partial_t\vct D,
  \quad
  \curl\vct H\sub d = \corr{\hat{T}(\partial_t\vct D)}
  = 0
  .
\end{align*}
Then,
in the second equation,
the displacement current $\partial_t\vct D$
certainly disappears and $\vct H\sub d$ vanishes.
But we should remember that it reapears in the first equation and contribute to $\vct H$.
We cannot erase the displacement current.

\section{Example of Helmholtz Decomposition of Current Distribution}


In \corr{Fig.\ref{fig:examples}},
the Helmholtz decomposition is shown for several current distributions.
All of these examples has been used to discuss displacement currents.
Each of them is briefly described below.
We have already mentioned
in Sec.~\ref{sec:biotsavart}
about the current elements in the first row.
In the case of point charge $q$ moving at
velocity $\vct v$,
we can set $\vct p = q\vct v$,
instead of $I\incr\vct l$.

\subsection{Semi-infinite linear current}

If a constant current $I$ is flowing along
the half line ($z\leq 0$) along the $z$-axis,
the charge must be accumulated at the origin
as
$Q(t) = It + Q(0)$, where $Q(0)$ is the charge at time $t=0$.

Electric flux density of Coulomb type and
the associated displacement current density
$
\partial_t\vct D(t, \vct r)=I\vct G(\vct r)
$
are generated.
We can derive the magnetic fields in two ways, i.e.,
with Biot-Savart's law and
Maxwell-Amp\`{e}re's formula.

Using cylindrical coordinates $(\rho,\phi,z)$, the
Biot-Savart law for an open path,
$L=\{\vct r'=z\vct e_z \mid -\infty<z\leq 0\}$,
we have
\begin{align*}
&
H_\phi(\vct r) =
H_\phi (\rho, z) =
\int_L I\dd z\vct e_z\times\vct G(\vct r-\vct r')
\\
&=\frac{I}{4\pi}
\int_{-\infty}^0 \frac{\rho\, \dd z'}{[(z-z')^2+\rho^2]^{3/2}}
=\frac{I}{4\pi\rho}\left(
1-\frac{z}{\sqrt{z^2+\rho^2}}\right)
.
\end{align*}

Secondly, to apply Maxwell-Amp\`{e}re's formula,
consider a spherical surface with radius $R=\sqrt{\varrho^2+z^2}$ centered
at the origin.
We have a latitude line $C$ defined by
$\theta=\tan^{-1}(\varrho/z)=\text{const}$.
Let $S_+$ ($S_-$) be the northern (southern) spherical crown
cut by $C$.
Note that $C = \partial S_+ = -\partial S_-$.
Applying the Maxwell-Amp\`{e}re formula (\ref{eq:maxwell-ampere}) for both surfaces, we have
\begin{align*}
\oint_C \vct H\cdot\dd\vct l
= \corr{\int_{S_+}\partial_t\vct D\cdot\dd\vct S}
= -\int_{S_-}(\vct J + \partial_t\vct D)\cdot\dd\vct S
.
\end{align*}
The left hand side is
$
2\pi\rho H_\phi(\rho, z)
$.
The middle side is evaluated as follows.
Since
the magnitude of the displacement current
on a sphere of radius $R$ is
$\partial_t D = I/(4\pi R^2)$,
it is perpendicular to the sphere, and
the area of $S_+$ is
$|S_+|=2\pi R^2(1-\cos\theta)$, we have
\begin{align*}
\partial_t D |S_+| = \frac{I}{2}(1-\cos\theta)
.
\end{align*}
Similary for the right hand side
using $|S_-|=2\pi R^2(1+\cos\theta)$
and
$\int_{S_-}\vct J\cdot\dd\vct S = \corr{-I}$, we have
\begin{align*}
I - \partial_t D |S_-| = I - \frac{I}{2}(1+\cos\theta)
.
\end{align*}
Thus,
the same result is obtained by using either of the surfaces:
\begin{align*}
  H_\phi(\rho, z) = \frac{I}{4\pi\rho}(1-\cos\theta)
  = \frac{I}{4\pi\rho}\left(1-\frac{z}{\sqrt{\rho^2+z^2}}\right)
  .
\end{align*}
This result agrees with the previous result by Bio-Savart's law.
In this method we had to consider the displacement currents
explicitly, otherwise,
the solution is not uniquely determined.

\subsection{Spherically symmetric current distribution}

To back up the false claim that displacement currents do not create magnetic fields,
the fallacious theory that "spherically symmetric currents do not create a magnetic field due to their symmetry" is developed.
Combining Biot-Savart's equation and symmetry for each part of the current,
it is attempted to to show that the magnetic field is zero.

As an example,
let's take a look at section 9.2
of Purcell's textbook \cite{purcell}.
Noting that a curl-free field like
displacement current density can be written as
a superposition of spherically symmetric Coulomb-type vector fields,
it continues,
\begin{quotation}
... the magnetic fields of any radial, symmetrical current
distribution, calculated via Biot-Savart, is zero.
To understand why, consider a radial line through a given line
through a given location.
At this location, the Biot-Savart's contributions from a pair of
points symmetrically located with respect to this line are equal and
opposite, as you can verify.
The contributions therefore cancel in pairs, yielding zero field
at the given location.
\end{quotation}
Here the fact is forgotten that Biot-Savart's law includes the effect of displacement current, which
in this case is spherically symmetric
but flows in the opposite direction.

In his famous series of text book \cite{feynman},
Feynman correctly explained the situation of spherically
symmetric current
using a model where a small sphere with radioactive
material is squirting out some charged particles.

Physically speaking, as shown in Fig.~\ref{fig:helmholz-decomposition},
spherically symmetric currents can be thought of
as a large number of semi-linear currents isotropically
combined at a single point.
In this case,
the displacement currents at the endpoints add up to exactly cancel the original currents.
The total current becomes zero in all places,
and therefore the magnetic field is also zero.
The magnetic field vanishes not owing to the cancellation of real currents.
In short,
\corr{if} the symmetric current (conducting or displacement)
is purely longitudinal,
the total current is zero.

\begin{figure}[t]
  \centering
  \includegraphics[scale=0.48]{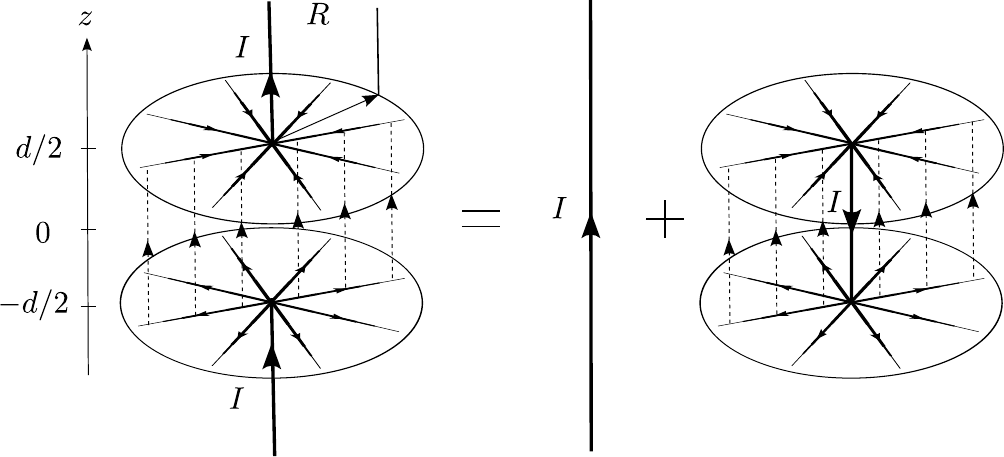}
  \caption{Charging capacitor.
  The spacing $d$ is exaggerated compared with the
  radius $R$.
  The displacement current is uniformly distributed.}
  \label{fig:charging-capacitor}
\end{figure}

\subsection{Charging capacitor}

The Charging capacitor problem is \corr{one} of the reasons why Maxwell came up with the idea of displacement current.
The debate continues over this model, as to whether the displacement current between the electrodes creates a magnetic field or not
\cite{roche1,jackson1,dahm}.

Surprisingly, miscalculations are found,
from time to time, in dealing with this problem.
For example,
Roche \cite{roche1}
made wrong calculation
in his historical survey paper on the controversy
over the reality of displacement current.
Jackson\cite{jackson1} pointed out the mistake
and complained that the paper is marred by imprecision, sloppy
notation, and downright mistakes.

For simplicity, let us assume an axisymmetric system around $z$-axis as shown in Fig.~\ref{fig:charging-capacitor}.
Let $R$ be the radius of the electrodes
and $d$ be the spacing.
The system is assumed to be charged
by a constant current $I$ through the conducting
wires.
Assuming that $d\ll R$, the electric
flux density between the electrodes
is spatially uniform and
given as $D_z(t)= It/\pi R^2$.

Since the current has no angular component, $J_\phi=0$, the magnetic field has only the
azimuthal component $H_\phi$.
The magnetic field due to the current $I$ on the straight conductor is
\begin{align}
  H\sur{(1)}_\phi(\rho, z)
= \frac{I}{2\pi\rho}
.
\label{eq:220}
\end{align}

The current $I$ in the wire $z<-d/2$
reaches the eletrode at $z=-d/2$ and
spreads radially and
the electrode is charged with a uniform charge surface density.
The linear density of this radial current
$K_\rho(\rho)$ is dependent on $\rho$
and can be determind by the charge conservation condition and the uniformity.
With the two-dimensional divergence formula we have
$\rho^{-1}(\dd/\dd \rho)[\rho K_\rho(\rho)] = \text{const}$.
Under the boundary conditions
$(2\pi\rho K_\rho)(0)=I$ and $K_\rho(R)=0$, the
differential equation can be solved;
\begin{align}
K_\rho(\rho) = \frac{I}{2\pi\rho}\left(1-\frac{\rho^2}{R^2}\right)
.
\label{eq:surface-current}
\end{align}
For the other electrode at $z=d/2$,
the surface current density is $-K_\rho(\rho)$.

The straight conductor has a gap
$-d/2 < z < d/2$ of length $d$.
The $H\sur{(1)}_\phi$ includes its contribution.
Therefore, we have to introduce a current of $-I$
flowing in this segment \cite{jackson1}.
(Roche \cite{roche1} missed this contribution,
which remains even in the limit, $d\rightarrow0$.)
If we connect this current element to the two surface currents,
the displacement currents from these two connection points become zero and
only those between the electrodes remain.

In order to apply
Amp\`{e}re-Maxwell's law,
we set up a circular loop,
$\rho=\text{const}$
between the electrodes ($-d/2<z<d/2$).
For $\rho\leq R$,
we have
$
  2\pi\rho H^{(2)}_\phi = -I + I{\rho^2}/{R^2}
$,
or
\begin{align*}
  H^{(2)}_\phi(\rho, z) = \frac{I}{2\pi\rho}\left(-1 + \frac{\rho^2}{R^2}\right)
  ,
\end{align*}
For $\rho>R$, we note $H^{(2)}_\phi = 0$.
By superposition, we have
the magnetic field of all position as
\begin{align}
  &H_\phi(\rho, z) = H^{(1)}_\phi(\rho, z) + H^{(2)}_\phi(\rho, z)
  \nonumber\\
  &\hspace{3ex}=
\begin{cases}
\displaystyle
\hspace{0.5em}\frac{I}{2\pi}\frac{\rho}{R^2} & (\rho\leq R,\,-d/2<z<d/2)
\\[2ex]
\displaystyle
\hspace{0.5em}\frac{I}{2\pi}\frac{1}{\rho} &
(\text{otherwise})
\end{cases}
.
\label{eq:charging-capacitor-field}
\end{align}
The difference between the magnetic fields inside and outside the electrodes
is equal to the surface current density
(\ref{eq:surface-current}):
\begin{align*}
H_\phi(\rho, d/2+0) - H_\phi(\rho, d/2-0) = K_\rho(\rho)
.
\end{align*}
This current splitting in Fig.~\ref{fig:charging-capacitor},
which follows the
rule (\ref{eq:proper-division}),
is a clever way to avoid displacement currents except those between these electrodes.

Many experiments on charging capacitor have been conducted \cite{cauwenberghe,carver,bartlett,scheler}.
Especially, the careful measurement by Bartlett
and Corle \cite{bartlett} seems very precise and
\corr{consistent} with Eq.~(\ref{eq:charging-capacitor-field}).
But the title of paper
``Measuring Maxwell's displacement current
inside a capacitor''
is misleading, because
what was measured was the magnetic field inside
the capacitor.
As authors admitted in the very last paragraph
(and later in \cite{bartlett-ajp}),
\begin{quote}
What we have shown, then, is that the Biot-Savart
law applies to open as well as to closed circuits.
One may write the differential form of this law as
$\dd\vct B= I\dd\vct l\times\vct R/R^3$,
without the usual caveat that only
the integral around
a closed loop is meaningful.
\end{quote}
The experiment does not
answer the question whether
the displacement current generate magnetic field or not
(or the question itself is meaningless).
In my opinion, it should be stressed
that the vortex of the magnetic field between the electrodes was actually measured.
The existence of such vortices cannot be
explained without displacement currents.
Further more it might be interesting to demonstrate that
when the wiring path to the capacitor is changed (e.g., to form a coil) so as
the magnetic field between the electrodes is disturbed, the uniform vortex is still maintained.

Magnetic fields generated by a \corr{capacitor} discharging
through the partially conducting spacer are
dicussed in confusion from time to time (leaky capacitor)
\cite{french}.
The field profile can better be understood in
terms of that for
a current element (see Fig.~\corr{\ref{fig:examples}}).

\section{Displacement Currents and Electromagnetic Waves}
Let us look at the relationship between displacement currents and electromagnetic waves utilizing the Helmholtz decomposition (\ref{eq:helmholtz-decomposition}).
From two of the Maxwell equations and the constitutive relation of vacuum
\begin{align}
& \pdfrac{\vct B}{t} = -\curl\vct E
,\quad
  \pdfrac{\vct D}{t} = \curl\vct H
\\
&  \vct D = \varepsilon_0\vct E
,\quad
  \vct H = \mu_0^{-1}\vct B
  ,
  \nonumber
\end{align}
we have the equations for
the plane waves propageted
in the $z$ direction
(assuming $\partial_x=\partial_y=0$ and $x$ polarization, i.e., $E_y=0$)
\begin{align*}
  \varepsilon_0\pdfrac{E_x}{z}=-\pdfrac{H_y}{t},
\quad {\mu_0\pdfrac{H_y}{z}}=-\pdfrac{E_x}{t}
.
\end{align*}
The d'Alembert solution to the hyperbolic
partial differential equations is
\begin{align*}
  E_x(t, z) = f(z - c_0 t) + g(z + c_0 t)
  .
\end{align*}
The $f$ and $g$ are arbitrary functions.
These waves are propagated at
the velocity $\pm c_0 = \pm\sqrt{\mu_0/\varepsilon_0}$.
If there were no displacement current term, then $c_0\rightarrow\infty$ and no wave solution existed.

So far, we assumed the case where the steady current condition is violated
($\diver\vct J\neq0$, i.e.
$\partial_t\vct D\neq0$).
We now relax the condition further and consider the case where
the magnetic field is also time-varying
($\partial_t\vct B\neq0$).

We separate Maxwell's equations into the longitudinal and transverse components.
Since the magnetic flux density satisfies $\diver\vct B=0$, we have
$\vct B \equiv \vct B\sub{T}$, or $\vct B\sub{L}=0$.
The equation of electromagnetic induction is
\begin{align}
\curl(\vct E\sub T + \vct E\sub L) =
\curl\vct E\sub{T}
= -\partial_t\vct B\sub{T}
.
\label{eq:induction}
\end{align}
For the electric flux density,
from $\diver(\vct D\sub T+\vct D\sub L)
= \diver\vct D\sub{L} = \varrho$,
the longitudinal component $\vct D\sub{L}$
is related to the charge density $\varrho$.
On the other hand, the transverse component $\vct D\sub{T} = \varepsilon_0\vct E\sub{T}$ is
related to the electromagnetic induction equation (\ref{eq:induction}).

Substituting
$\vct H=\vct H\sub{T}=\mu_0^{-1}\vct B\sub{T}$,
into Maxwell-Amp\`{e}re's equation, we have
\begin{align*}
\curl\vct H\sub{T} &=
\vct J\sub{T} + \partial_t\vct D\sub{T}
 +
(
\vct J\sub{L} + \partial_t\vct D\sub{L}
)
.
\end{align*}
Since the left-hand side is a transverse field (divergence-free), as in the case of a stationary field,
the longitudinal components cancel each other;
$\vct J\sub{L} + \partial_t \vct D\sub{L} \equiv 0$.
This can be regarded as the role of the longitudinal component of the displacement current.
In summary, we have
\begin{align}
&\curl\vct E\sub{T} = -\partial_t\vct B\sub{T}
,\quad
\curl\vct H\sub{T} = \vct J\sub{T} + \partial_t\vct D\sub{T},
\\
&\vct B\sub{L}=0,
\quad
\diver\vct D\sub{L}=\varrho,
\quad
\vct J\sub{L}+\partial_t\vct D\sub{L}=0
.
\nonumber
\end{align}
Since there is no divergence for both fields
$\vct J\sub{T}$ and $\partial_t\vct D\sub{T}$,
they can be splitted properly and
the magnetic field created by each of them can be defined.
By the coupling of the first two equations due to the transverse displacement currents,
hybridization of the transverse components of the electric and magnetic fields, i.e.,
electromagnetic wave modes are enabled.

In particular,
for $\vct J\sub{T}=0$, a free solution exists.
Since the transverse displacement current $\partial_t\vct D\sub{T}$ is
not bounded by the real current,
the electromagnetic wave can propagate far from the source.
In fact the dependence of the longitudinal component of the displacement current on the distance from the source is at most $1/R^2$,
that of the electromagnetic wave, i.e., the transverse component, is $1/R$.

Sometimes the Jeffimenko equation,
which is a dynamical extension of Coulombs's law and Biot-Savart's law, is used to rule out the contribution
of displacement currents in quasi-static cases
\cite{griffiths}.
But arguments in this direction
only complicate things and do not seem useful.

\section{Summary}
The controversy
over the meaning of displacement currents tends to get lost in the choice between creating a magnetic field or not.
Such confusion arises from
the following facts.
\begin{itemize}
\item
In the case of non-stationary currents, neither
magnetic field created by conduction current nor
that created by displacement current can be defined.
\item
In solving Maxwell-Amp\`{e}re's equation by superposition,
the right-hand side
cannot be divided arbitrary ignoring
the inseparability of the displacement current and the current.
\item
The effect of displacement current is automatically incorporated in the magnetic field calculated by Biot-Savart's law.
\end{itemize}

The existence of displacement current is subtle and elusive, and it was only discovered through Maxwell's deep insight.
He emphasized the importance of the unity of the current and the displacement current, and defined the sum of them as
total current $\vct J\sub{tot}$.
He wrote down the fundamental equations of the electromagnetic field using the total current $\vct J\sub{tot}$.

As we have seen,
each of the longitudinal and transverse components of
displacement currents plays a different role.
Although the former was the initial impetus for
the introduction of displacement currents,
it plays a shadow role
to counteract the longitudinal component of the current.
On the other hand,
the latter
plays a prominent role
in propagating electromagnetic waves

While attempting to distinguish the roles of current and displacement current,
researchers have fallen for the trap of Bio-Savart's equation.
In particular, the statement that the displacement current does not produce a magnetic field may
leads to underestimation of displacement currents,
or even to lost sight of its essential role in electromagnetic waves.


\end{document}